\documentclass[12pt]{article}

\usepackage{enumitem}
\usepackage{graphicx}
\usepackage{float}
\usepackage{cite}
\usepackage{amsfonts}
\usepackage{amssymb}
\usepackage{amsmath}
\usepackage{xcolor}

\def\gtwid{\mathrel{\raise.3ex\hbox{$>$\kern-.75em\lower1ex\hbox{$\sim$}}}}
\def\ltwid{\mathrel{\raise.3ex\hbox{$<$\kern-.75em\lower1ex\hbox{$\sim$}}}}
\def\square{\kern1pt\vbox{\hrule height 1.2pt\hbox{\vrule width 1.2pt\hskip 3pt
  \vbox{\vskip 6pt}\hskip 3pt\vrule width 0.6pt}\hrule height 0.6pt}\kern1pt}

\begin{document}

\begin{titlepage}

\begin{flushright}
CCTP-2026-01 \\
ITCP/2026/01 \\
UFIFT-QG-26-02
\end{flushright}

\vskip 1cm

\begin{center}
{\bf Quantum Cosmology in Accelerating Spacetimes II}
\end{center}

\vskip 1cm

\begin{center}
S. P. Miao$^{1\star}$, N. C. Tsamis$^{2\dagger}$ and 
R. P. Woodard$^{3\ddagger}$
\end{center}

\vskip 0.5cm

\begin{center}
\it{$^{1}$ Department of Physics, National Cheng Kung University, \\
No. 1 University Road, Tainan City 70101, TAIWAN}
\end{center}

\begin{center}
\it{$^{2}$ Institute of Theoretical Physics \& Computational Physics, \\
Department of Physics, University of Crete, \\
GR-700 13 Heraklion, HELLAS}
\end{center}

\begin{center}
\it{$^{3}$ Department of Physics, University of Florida,\\
Gainesville, FL 32611, UNITED STATES}
\end{center}

\vspace{0cm}

\begin{center}
ABSTRACT
\end{center}
This paper is a sequel in which we further analyze 
the recently derived quantum gravity equations which
apply in accelerating cosmological spacetimes and
whose solutions should be equivalent to all order 
re-summations of the perturbative leading logarithms
that appear. In particular we study their implications 
concerning the primordial tensor power spectrum and 
the gravitational force due to a test source.
\begin{flushleft}
\end{flushleft}

\vskip 0.5cm

\begin{flushleft}
$^{\star}$ e-mail: spmiao5@mail.ncku.edu.tw \\
$^{\dagger}$ e-mail: tsamis@physics.uoc.gr \\
$^{\ddagger}$ e-mail: woodard@phys.ufl.edu
\end{flushleft}

\end{titlepage}

\section{Prologue}

A standard example of an accelerating spacetime which
seems to be of physical significance during early cosmological 
evolution is the primordial inflationary epoch defined by 
$H > 0$ with $0 \leq \epsilon < 1$ \cite{Geshnizjani:2011dk}.
Any cosmological geometry is characterized by a scale factor 
$a(t)$ and its two first time derivatives, the Hubble parameter 
$H(t)$ and the 1st slow roll parameter $\epsilon(t)$:
\footnote{It is more often than not convenient to employ 
conformal instead of co-moving coordinates:
$ds^2 \!=\! 
-dt^2 + a^2(t) \, d{\mathbf{x}} \cdot d{\mathbf{x}} 
= 
a^2(\eta) \big[\! -d\eta^2 + a^2(\eta) \, 
d{\mathbf{x}} \cdot d{\mathbf{x}} \big]$, with
$t$ the co-moving time and $\eta$ the conformal
time.}
\begin{equation}
ds^2 = - dt^2 + a^2(t) \, d{\mathbf{x}} \cdot d{\mathbf{x}} 
\qquad \Longrightarrow \qquad 
H(t) \equiv \frac{\dot{a}}{a} 
\quad, \quad 
\epsilon(t) \equiv - \frac{\dot{H}}{H^2} 
\; . \label{geometry}
\end{equation}
However, there is no agreement on the cause of  primordial 
inflation \cite{Ijjas:2013vea,Guth:2013sya,Linde:2014nna}, 
nor is it understood how the parameters that characterize 
the current universe assumed their sometimes unexpected
values. A reasonable conjecture is that the answer may 
involve quantum gravitational corrections, which grew to 
become nonperturbatively strong during a prolonged era 
of inflation \cite{Tsamis:1996qm, Tsamis:2011ep}; after 
all, it is gravitation which plays the dominant role in 
shaping cosmological evolution.
\footnote{It goes without saying that the study of quantum 
physics in de Sitter spacetime, the maximally symmetric 
example of an accelerating spacetime, already has a long 
history \cite{Grishchuk:1977zz,
Myhrvold:1983hx,
Ford:1984hs,
Allen:1985wd, 
Antoniadis:1986sb, 
Allen:1986tt,
Floratos:1987ek, 
Higuchi:1991tm,
Tsamis:1992xa,
Dolgov:1994cq,
Tsamis:2005hd,
Polyakov:2007mm,
Giddings:2007nu,
Perez-Nadal:2007yxe,
Burgess:2010dd,
Polyakov:2012uc,
Marolf:2012kh, 
Anderson:2013ila, 
Frob:2013ht,
Anninos:2014lwa,
Frob:2014zka,
Dvali:2014gua, 
Burgess:2015ajz, 
Brandenberger:2018fdd,
Baumgart:2019clc,
Brahma:2021mng,
Colas:2022hlq,
Cable:2023gdz, 
Burgess:2024eng,
Anninos:2024fty,
Brahma:2024yor,
Sloth:2025nan,
Ansari:2025nng,
Ahmadi:2025oon,
Kaplanek:2025moq,
Prokopec:2025jrd}.} 

Although quantum gravity is not perturbatively renormalizable, 
it can still be used reliably as a quantum low-energy effective 
field theory in the generic sense of Weinberg \cite{DHoker:1994rdl} 
as, for instance, done by Donoghue 
\cite{Donoghue:1993eb,Donoghue:1994dn,Donoghue:2017ovt}.
Quantum gravity was much stronger during primordial inflation 
than today because the relevant loop-counting parameter 
$\tfrac{\hbar G H^2(t)}{c^5}$ -- which is only about $10^{-122}$ 
currently -- may have been as large as $10^{-11}$ towards the end 
of inflation and even bigger before. While this is small enough 
to make perturbation theory valid, it is large enough to produce 
detectable effects \cite{Starobinsky:1979ty,Mukhanov:1981xt}. 

These effects survive to the current epoch because light, 
minimally coupled scalars and gravitons fossilize when 
their physical wave numbers $\tfrac{k}{a(t)}$ fall below 
the inverse Hubble length $\tfrac{H(t)}{c}$ 
\cite{Mukhanov:1990me}. Since this horizon crossing continues 
as long as accelerated expansion persists, loop corrections 
that would be constant in flat space tend to grow with time, 
and sometimes also with space. For instance, the vacuum 
polarization in de Sitter spacetime \cite{Leonard:2013xsa} 
from a loop of gravitons changes the electric field strength 
$F_{0i}(t,\bf{x})$ of plane wave photons \cite{Wang:2014tza} 
and the Coulomb $\Phi(t, r)$ potential \cite{Glavan:2013jca} 
to:
\footnote{The $G \, r^{-2}\!$ correction in (\ref{Coulomb}) 
is the cosmological descendant of an effect long known from 
flat space background \cite{Radkowski:1970ovx}.}
\begin{eqnarray}
F_{0i}(t,{\bf x}) &\!\!\! = \!\!\!& 
F^{\rm tree}_{0i}(t,{\bf x}) \, \Bigl\{
1 + \tfrac{2 \hbar G H^2}{\pi c^5} \ln[a(t)] + \dots \Bigr\} 
\; , \label{photons} \\
\Phi(t,r) &\!\!\! = \!\!\!& 
\tfrac{Q}{4\pi \varepsilon_0 \, a(t) \, r} \, \Bigl\{
1 + \tfrac{2 \hbar G}{3 \pi c^3 a^2(t) \, r^2} 
+ \tfrac{2 \hbar G H^2}{\pi c^5} \ln[ \tfrac{a(t) H r}{c}] 
+ \dots \Bigr\} 
\; . \label{Coulomb}
\end{eqnarray}
The large logarithmic terms proportional to $G H^2$ in
(\ref{photons}-\ref{Coulomb}) are {\it new} effects that 
derive from inflationary gravitons. 
\footnote{A number of similar logarithmic terms in de Sitter
background can be found in \cite{Miao:2005am,Miao:2006gj,
Tan:2021ibs,Tan:2021lza,Glavan:2021adm,Tan:2022xpn}.} 

By studying non-linear sigma models \cite{Miao:2021gic}, 
and proceeding to scalar loop corrections to gravity 
\cite{Miao:2024nsz}, it can be shown that the large 
logarithmic terms can be {\it resummed} by combining 
a variant of Starobinsky's original stochastic formalism 
\cite{Starobinsky:1986fx} with a variant of the 
renormalization group. For instance, scalars on de Sitter 
spacetime change the electric components of the Weyl tensor 
$C_{0i0j}$ for plane wave gravitons, and the Newtonian 
potential $\Psi$ to \cite{Miao:2024nsz,Miao:2024atw}:
\begin{equation}
C_{0i0j} = 
C^{\rm tree}_{0i0j} \, \Bigl\{ 
1 - \tfrac{3 \hbar G H^2}{10 \pi c^5} \ln[a] + \dots\Bigr\} 
\; \longrightarrow \; 
C^{\rm tree}_{0i0j} \!\times\! 
[ a(t)]^{-\frac{3 \hbar G H^2}{10 \pi c^5}} 
\, , \qquad\qquad \label{Weyl} 
\end{equation}
\vspace{-0.5cm}
\begin{equation}
\Psi = 
\tfrac{G M}{a r} \, \Bigl\{ 
1 + \tfrac{\hbar G}{20 \pi c^3 a^2 r^2} 
- \tfrac{3 \hbar G H^2}{10 \pi c^5} \ln[\tfrac{a H r}{c}] 
+ \dots \Bigr\} 
\; \longrightarrow \; 
\tfrac{G M}{a r} \!\times\! 
[ \tfrac{a(t) H r}{c}]^{-\frac{3 \hbar G H^2}{10 \pi c^5}}
, \label{Newtonscalar}
\end{equation}
where the results on the far right represent the 
non-perturbative resummations. 

Most recently, a set of rather simple stochastic field 
equations has been derived for pure quantum gravity in 
a fixed gauge \cite{Miao:2024shs,Miao:2025gzm}. Unlike 
previous perturbative loop computations, these equations 
can, in principle, be solved {\it non-perturbatively} as 
long as the expansion rate changes slowly. On one hand, 
this set of equations in its perturbative correspondence 
limit predicts changes to the expansion rate 
\cite{Miao:2025bmd} and -- as we shall see in the present 
study -- to the tensor power spectrum and the gravitational 
force, all of which grow non-perurbatively strong. On the 
other hand however, this set of equations still needs to 
complete demonstrating its consistency.


\section{The appropriate gravitational equations}

The general $D$-dimensional gravitational effective theory 
satisfying general coordinate invariance is an infinite 
series of terms with increasing canonical dimensionality. 
For cosmology the two lowest terms suffice:
\footnote{Hellenic indices take on spacetime values
while Latin indices take on space values. Our metric
tensor $g_{\mu\nu}$ has spacelike signature
$( - \, + \, + \, +)$ and our curvature tensor equals
$R^{\alpha}_{~ \beta \mu \nu} \equiv
\Gamma^{\alpha}_{~ \nu \beta , \mu} +
\Gamma^{\alpha}_{~ \mu \rho} \,
\Gamma^{\rho}_{~ \nu \beta} -
(\mu \leftrightarrow \nu)$.}
\begin{equation}
{\mathcal L}_{inv} =
\frac{1}{\kappa^2} \Big[ \!-\! (D \!-\! 2) \Lambda + R \, \Big] \sqrt{-g}
\; , \label{Linv}
\end{equation}
where the two parameters are Newton's constant $G$ and the 
cosmological constant $\Lambda$:
\footnote{$\Lambda$ is taken to be ``large'' and positive. 
Here ``large'' means a $\Lambda$ corresponding to a scale
$M$ which can be as high as $10^{18} \, GeV$.} 
\begin{equation}
\kappa^2 \equiv 16 \pi G
\quad , \quad
\Lambda \equiv (D \!-\! 1) H^2
\; . \label{parameters}
\end{equation}
The full equations emanating from (\ref{Linv}) are:
\begin{equation}
R_{\mu\nu} - \frac12 R \, g_{\mu\nu} 
+ \frac12 (D \!-\! 2)(D \!-\! 1) H^2 g_{\mu\nu} = 0
\; . \label{eom}
\end{equation}
In terms of the full metric $g_{\mu\nu}$, the conformally
rescaled metric ${\widetilde g}_{\mu\nu}$ and the graviton 
field $h_{\mu\nu}$ are defined thusly:
\begin{equation}
g_{\mu\nu} \equiv 
a^2 {\widetilde g}_{\mu\nu} \equiv
a^2 \big[ \eta_{\mu\nu} + \kappa h_{\mu\nu} \big] 
\; . \label{metrics}
\end{equation}

The graviton field variables are defined as follows:
\begin{equation}
\kappa \, h_{\alpha\beta}
\equiv 
A_{\alpha\beta}
+ u_{(\alpha} B_{\beta)}
+ ( u_{\alpha} u_{\beta} \!+\! \overline{\gamma}_{\alpha\beta} ) C
\quad , \quad
u^{\alpha} A_{\alpha\beta} = u^{\alpha} B_{\alpha} = 0
\; , \label{ABC}
\end{equation}
where $A_{\alpha\beta}$ contains the dynamical degrees 
of freedom, $B_{\alpha}$ is canonically associated with 
the momentum constraints of general relativity and physically 
represents the strictly general relativistic potentials, 
and $C$ is canonically associated with the Hamiltonian 
constraint of general relativity and physically represents 
the Newtonian potential; in addition, $C$ is the constrained 
field variable in which any dynamical changes of the 
geometrical background are imprinted.

The geometrical variables are those of the Arnowitt-Deser-Misner
(ADM) spatial-temporal decomposition of the conformally re-scaled 
metric \cite{Arnowitt:1962hi}:
\begin{eqnarray}
d{\widetilde s}^{\,\,\! 2} 
& \!\!\!=\!\!\! & 
- N^2 d\eta^2 + \gamma_{ij} (dx^i - N^i d\eta) (dx^j - N^j d\eta)
\; , \label{3+1a} \\
& \!\!\!=\!\!\! & 
(- N^2 \!+\! \gamma_{ij} N^i N^j) \, d\eta^2
- 2 \gamma_{ij} N^j d\eta \, dx^i + \gamma_{ij} dx^i dx^j
\; , \label{3+1b}
\end{eqnarray}
where $N$ is the lapse function, $N^i$ is the shift function, 
and $\gamma_{ij}$ is the spatial metric.
\footnote{The four-vector $u_{\mu}$ -- the "temporal part" --
is defined in the Appendix, where the ADM decomposition basics 
can be viewed.}

The field variables $A_{ij}, B_i, C$ are related to the 
geometrical variables $\gamma_{ij}, N^i, N$ thusly:
\begin{equation}
N^2 = \frac{1}{1 + C + \tfrac14 B_i B_i}
\;\; , \;\;
N^i = \frac{\tfrac12 B_i}{\sqrt{1 + C + \tfrac14 B_i B_i}}
\;\; , \;\;
\gamma_{ij} = \frac{\delta_{ij} + A_{ij}}{1 - C}
\; . \label{relation} 
\end{equation}

Finally, the gauge fixing Lagrangian term employed equals:
\begin{equation}
{\widetilde{\mathcal L}}_{GF} =
- \tfrac12 a^{D-2} \sqrt{-{\widetilde g}} \,
{\widetilde g}^{\mu\nu} \, {\widetilde F}_{\mu} \,
{\widetilde F}_{\nu}
\; , \label{Lgf} \\
\end{equation}
where the gauge condition is:
\begin{equation}
{\widetilde F}_{\mu} =
{\widetilde g}^{\rho\sigma} \,
\Big[ h_{\mu\rho, \sigma} - \tfrac12 h_{\rho\sigma, \mu} \,
- (D-2) a {\widetilde H} h_{\mu\rho} u_{\sigma} \Big]
\quad , \quad
{\widetilde H} \equiv \frac{H}{N}
\; . \qquad \label{F}
\end{equation}
We should like to appropiately reduce the gravitational
field equations (\ref{eom}) in a way which still retains
the large logarithmic terms that appear in perturbation
theory for accelerating spacetimes.\footnote{The analysis
and the results that follow have been obtained for de Sitter, 
the prototypical physical example of an accelerating spacetime 
and one in which explicit calculations are feasible.} 
This was accomplished sequentially in 
\cite{Miao:2024shs,Miao:2025gzm,Miao:2025bmd} where all the
details can be found; a hopefully adequate summary follows. 

Dimensionally regularized, and fully renormalized perturbative
computations can exhibit secular effects characterized by 
powers of $ln(a)$ due to two sources:
\\ [3pt]
{\it (i)} The logarithmic part of the propagator 
{\it (stochastic logs)} which can be schematically
expressed as:
\footnote{DeWitt and Brehme termed this part of the graviton
propagator as the ``tail'' part \cite{DeWitt:1960fc}.}
\begin{equation}
i \Delta \sim \tfrac{1}{x^2} - \ln(x^2)
\; , \label{TailLogs}
\end{equation}
Quantum fields which possess such ``tails'' are known as 
{\it active}; all other fields are termed {\it passive}.
\\ [3pt]
{\it (ii)} The incomplete cancellation between primitive 
divergences and counterterms {\it (renormalization induced logs)}
which can be schematically expressed as:
\begin{equation}
\tfrac{H^{D-4}}{D-4} - \tfrac{(\mu a)^{D-4}}{D-4}
=
- \ln\left( \tfrac{\mu a}{H} \right) + \mathcal{O}(D-4)
\; . \label{UVlogs}
\end{equation}
Any type of quantum field can contribute this kind of 
logarithm.

{\bf -} {\it Case I:} When interactions contain only 
undifferentiated active fields, the leading logarithms 
derive entirely from (\ref{TailLogs}) and can be recovered 
by replacing the full quantum field with a stochastic 
random field described by Starobinsky's formalism
\cite{Starobinsky:1986fx,Starobinsky:1994bd}. 
\footnote{The amazing fact that (ultraviolet divergent) 
correlators of the full quantum field $\phi(x)$ agree, 
at leading logarithm order, with those of the stochastic 
random field was proved in \cite{Tsamis:2005hd}.}

{\bf -} {\it Case II:} When undifferentiated active fields 
interact with passive fields, the leading logarithms again 
derive from the tail (\ref{TailLogs}). However, it is not 
correct to stochastically truncate passive fields because 
their effects derive as much from the ultraviolet as from 
the infrared, and from their full mode functions. The leading 
logarithms are instead recovered by integrating out the 
passive fields in a constant active field background, which 
produces a standard effective potential through the dependence 
of the passive field mass on the active field. One then
applies Starobinsky's formalism to the effective potential. 
\footnote{Examples include a fermion Yukawa-coupled to a 
massless minimally coupled scalar \cite{Miao:2006pn} and 
massless minimally couled  scalar electrodynamics 
\cite{Prokopec:2007ak, Prokopec:2008gw}.}

{\bf -} {\it Case III:} When interactions involve 
differentiated active fields, one generally encounters both 
kinds of logarithms.\footnote{Examples include non-linear 
sigma models \cite{Miao:2021gic}, scalar corrections to gravity 
\cite{Miao:2024nsz,Miao:2024atw} and pure gravity 
\cite{Miao:2024shs,Miao:2025gzm}.}
Stochastic logarithms (\ref{TailLogs}) are captured by integrating 
out the differentiated fields in the presence of a constant active 
field background. This produces new kinds of effective potentials 
based on how the active background affects field strengths or 
geometries. Renormalization induced logarithms (\ref{UVlogs}) 
can be treated using a variant of the renormalization group by 
exploiting the close relation between the scale factor $a(t)$
and the regularization scale $\mu$, which is evident in 
expression (\ref{UVlogs}). The variation consists of expressing 
Bogoliubov-Parasiuk-Hepp-Zimmermann (BPHZ) counterterms as a 
higher derivative of the quantum field, plus renormalizations 
of bare parameters, for example:
\begin{equation}
R^2 = (R - D \Lambda)^2 + 2 D \Lambda [R - (D\!-\!2) \Lambda)]
+ D (D \!-\! 4) \Lambda^2 \; . \label{Eddington}
\end{equation}
While the factors of $\ln(a)$ from the first term are suppressed 
by powers of $a$, those from the second term can be summed using 
the Callan-Symanzik equation \cite{Miao:2024nsz}.

The result of applying {\it (i)} the rules for isolating 
the stochastic leading logarithms of pure quantum gravity 
and, {\it (ii)} the cosmological principle, is the following 
remarkably simple form for the leading logarithm (LLOG) 
gravitational equations in accelerating constant background 
cosmological spacetimes in $D=4$ \cite{Miao:2025bmd}:
\begin{eqnarray}
{\dot A}_{ij} \!-\! \mathcal{\dot A}_{ij} 
& \!\!\!=\!\!\! & 
\tfrac43 \frac{\kappa^2 \widetilde{H}^3}{8 \pi^2} 
\, \gamma_{ij} 
\Big[ 2 \!-\! \gamma^{rs} A_{rs} \Big] 
\; , \label{Afinal3} \\
B_i 
& \!\!\!=\!\!\! &
0
\; , \label{Bfinal3} \\
C 
& \!\!\!=\!\!\! & 
\frac{\kappa^2 \widetilde{H}^2}{8 \pi^2} 
\Big[ \!-\! 1 \!+\! 2 \gamma^{ij} A_{ij} \Bigr] 
\; . \label{Cfinal3}
\end{eqnarray}
The set of operator LLOG equations (\ref{Afinal3}-\ref{Cfinal3})
-- one differential and two algebraic -- contain the leading 
logarithms from all orders of perturbation theory and can be 
seen to be a {\it huge} simplification of the original field 
equations of pure gravity in the gauge (\ref{F}).\footnote{It
is worth mentioning at this stage that classifying according to
the strength of the logarithmic terms is a well-defined procedure.
Because pure gravity is a two parameter theory -- $G$ \& $\Lambda$
-- and the dimensionless coupling constant is the product
$G \Lambda \, \ln(a)$, we can classify diagrams in a well-defined 
way by $G \Lambda$ or by $\ln(a)$.}

The operator equations (\ref{Afinal3}-\ref{Cfinal3}), 
have the ability to obtain the complete (non-perturbative) 
LLOG time evolution of a physical quantity; this is still 
a highly non-trivial project to say the least but now it 
is a project which has perhaps become a rather realistic 
endeavour. Moreover, there is another way to use this set
of operator equations by considering their perturbative 
order by order limit; doing so permits us to easily determine 
the leading logarithm result from loop calculations which 
would normally take a long long time to conclude.
\footnote{Starobinsky exploited both ways to obtain both 
non-perturbative and perturbative results in the late time 
limit for his simple self-interacting scalar field model 
\cite{Starobinsky:1986fx,Tsamis:2005hd}.}

\section{Iterating the $A_{ij}$ equation}

It seems to be clear that the above set of LLOG equations 
(\ref{Afinal3}-\ref{Cfinal3}) must eventually be solved 
to simultaneously obtain the leading logarithms from all 
orders of perturbation theory. Nonetheless, in the meantime
an important and highly non-trivial check on these equations 
is to consider their correspondence limit with perturbation
theory. In this direction, we already analyzed their 
perturbative prediction concerning the expansion rate in 
\cite{Miao:2025bmd}. Here we shall do the same for the mode
functions of the graviton -- from which the tensor power 
spectrum follows -- and for the gravitational force.
The results {\it must} agree with their fully regularized
and renormalized analogues from perturbation theory. 

In the system of LLOG equations, the constrained fields $B_i$ 
\& $C$ obey the algebraic equations (\ref{Bfinal3}-\ref{Cfinal3}) 
and start very small (or zero) and grow very slowly (or stay zero),
as they {\it respond} to the growth of the dynamical field $A_{ij}$ 
which obeys the differential equation (\ref{Afinal3}).

To perturbatively solve (\ref{Afinal3}) we first assume we have
already made a finite renormalization of the cosmological constant 
$\Lambda$ to absorb the constant term, so that:
\begin{equation}
\dot{A}_{ij} \simeq 
\dot{\mathcal{A}}_{ij} - \tfrac{\kappa^2 H^3}{6 \pi^2}
\gamma_{ij} \gamma^{k\ell} A_{k \ell} 
\; . \label{Aeqn}
\end{equation}
It is convenient to define the dimensionless parameters:
\begin{equation}
\lambda \equiv \tfrac{\kappa^2 H^2}{3 \pi^2} 
\qquad , \qquad 
T \equiv H t 
\; . \label{defs}
\end{equation}
because it allows us to express (\ref{Aeqn}) in a simple, 
dimensionless form:
\footnote{Here a prime denotes differentiation by $T$.}
\begin{equation}
A'_{ij} \simeq 
\mathcal{A}'_{ij} 
- \tfrac{\lambda}{2} \gamma_{ij} \gamma^{k\ell} A_{k\ell} 
\; . \label{NewAeqn}
\end{equation}

In developing a perturbative solution, the coincidence limit
of $\mathcal{A}_{ij}$ plays a central role:
\footnote{The result (\ref{Avev}) is derived for de Sitter 
spacetime \cite{Tsamis:1992xa} and is valid perturbatively 
about this background. In the Appendix some basic properties 
are displayed.}
\begin{equation}
\langle \mathcal{A}_{ij} \mathcal{A}_{k\ell} \rangle = 
\tfrac{\kappa^2 H^2 T}{4 \pi^2} \, [\mbox{}_{ij} T^A_{k\ell}] = 
\tfrac34 \lambda \, T \, [\mbox{}_{ij} T^A_{k\ell}] 
\; . \label{Avev}
\end{equation}

In the perturbative regime we operate, the exact relation 
(\ref{relation}) for $\gamma_{ij}$ becomes:
\begin{equation}
\gamma_{ij} = \frac{\delta_{ij} + A_{ij}}{1 - C}
\; \longrightarrow \;
\delta_{ij} + A_{ij}
\; , \label{gammaijpert}
\end{equation}
because $C$ -- which initially is zero and varies between
+1 and -1 -- perturbatively stays close to zero. Consequently,
(\ref{NewAeqn}) becomes:
\begin{equation}
A'_{ij} \simeq 
\mathcal{A}'_{ij} 
- \tfrac{\lambda}{2} [\delta_{ij} + A_{ij}] 
[A_{kk} - A_{k\ell} A_{\ell k} + A_{k\ell} A_{\ell m} A_{m k} - \dots ] 
\; , \label{pertAeqn}
\end{equation}
and we shall solve it to the requisite order.
\\ [3pt]
{\bf -} $\mathcal{O}(\lambda)$: 
At order $\lambda$ equation (\ref{pertAeqn}) reads,
\begin{equation}
{A^{(1)}}'_{ij} = \mathcal{A}_{ij}' 
\qquad \Longrightarrow \qquad 
A^{(1)}_{ij} = \mathcal{A}_{ij} 
\; . \label{A1}
\end{equation}
{\bf -} $\mathcal{O}(\lambda^2)$:
At order $\lambda^2$ equation (\ref{pertAeqn}) is,
\begin{equation}
{A^{(2)}}'_{ij} = 
-\tfrac{\lambda}{2} \delta_{ij} \, \mathcal{A}_{kk} 
\qquad \Longrightarrow \qquad
A^{(2)}_{ij} = 
-\tfrac{\lambda}{2} \, I[\mathcal{A}_{kk}] \delta_{ij} 
\; , \label{A2}
\end{equation}
where here and henceforth, the symbol $I[f]$ stands for 
the integral of the function $f(T)$ with respect to $T$:
\begin{equation}
I[f] \equiv \int_0^{T} \!\!\! dT' \, f(T') 
\; . \label{Idef}
\end{equation}
{\bf -} $\mathcal{O}(\lambda^3)$:
At order $\lambda^3$ equation (\ref{pertAeqn}) equals,
\begin{equation}
{A^{(3)}}'_{ij} = 
-\tfrac{\lambda}{2} \delta_{ij} A^{(2)}_{kk} 
- \tfrac{\lambda}{2} A^{(1)}_{ij} A^{(1)}_{kk} 
+ \tfrac{\lambda}{2} A^{(1)}_{k\ell} A^{(1)}_{\ell k} \delta_{ij} 
\; . \label{A3eqn}
\end{equation}
The solution to (\ref{A3eqn}) is:
\begin{equation}
A^{(3)}_{ij} = 
\Bigl\{ \tfrac34 \lambda^2 I^2[\mathcal{A}_{kk}] 
+ \tfrac{\lambda}{2} \, I[\mathcal{A}_{k\ell} \mathcal{A}_{\ell k}] 
\Bigr\} \delta_{ij} 
- \tfrac{\lambda}{2} \, I[\mathcal{A}_{kk} \mathcal{A}_{ij} ] 
\; . \label{A3}
\end{equation}
{\bf -} $\mathcal{O}(\lambda^4)$:
At order $\lambda^4$ equation (\ref{pertAeqn}) equals,
\begin{eqnarray}
\lefteqn{{A^{(4)}}'_{ij} = 
-\tfrac{\lambda}{2} \Bigl[ 
A^{(3)}_{kk} - 2 A^{(2)}_{k\ell} A^{(1)}_{\ell k} 
+ A^{(1)}_{k\ell} A^{(1)}_{\ell m} A^{(1)}_{mk} \Bigr] \delta_{ij} }
\nonumber \\
& & \hspace{1.9cm} 
- \tfrac{\lambda}{2} A^{(1)}_{ij} \Bigl[ 
A^{(2)}_{kk} - A^{(1)}_{k\ell} A^{(1)}_{\ell k} \Bigr] 
- \tfrac{\lambda}{2} A^{(2)}_{ij} A^{(1)}_{kk} 
\; . \label{A4eqn}
\end{eqnarray}
The solution to (\ref{A4eqn}) is:
\begin{eqnarray}
\lefteqn{A^{(4)}_{ij} = 
-\Bigl\{ \tfrac{9}{8} \lambda^3 I^3[\mathcal{A}_{kk}] 
+ \tfrac{\lambda}{4} \, I^2[3 \mathcal{A}_{k\ell} \mathcal{A}_{\ell k} 
  - \mathcal{A}_{kk} \mathcal{A}_{\ell\ell} ] 
+ \tfrac{\lambda^2}{8} ( I[\mathcal{A}_{kk}])^2 } 
\nonumber \\
& & \hspace{1.1cm} 
+ \tfrac{\lambda}{2} \, I[\mathcal{A}_{k\ell} \mathcal{A}_{\ell m} 
  \mathcal{A}_{mk}] \Bigr\} \delta_{ij} 
+ \tfrac34 \lambda^2 I\Bigl[ I[\mathcal{A}_{kk}] \mathcal{A}_{ij}\Bigr] 
+ \tfrac{\lambda}{2} \, I[\mathcal{A}_{k\ell} \mathcal{A}_{\ell k} 
\mathcal{A}_{ij}] \; . \qquad \label{A4}
\end{eqnarray}
We need not go to any higher orders. 

Finally, employing expression (\ref{Avev}) in equation 
(\ref{pertAeqn}), it follows that the vacuum 
expectation value of $A_{ij}$ equals:
\footnote{In (\ref{pertAeqn}), the lowest order contributing
terms are $+\frac{\lambda}{2} \delta_{ij} A_{k\ell} A_{\ell k}$
and $-\frac{\lambda}{2} A_{ij} A_{kk}$ upon effecting the lowest
order substitution $A_{ij} \rightarrow \mathcal{A}_{ij}$.}
\begin{equation}
\langle A_{ij} \rangle = 
\tfrac{15}{8} \lambda^2 T^2 \delta_{ij} + \mathcal{O}(\lambda^3 T^3) 
\; , \label{pertAvev}
\end{equation}
which implies that the first non-trivial secular contributions
appear at 2-loop order.

\section{Physical Consequences}

As we shall see, to compute corrections to the graviton 
mode functions and to the gravitational force, the expansion 
of $A_{ij}$ in powers of $\mathcal{A}_{ij}$
(\ref{A1},\ref{A2},\ref{A3},\ref{A4}) is involved in both
cases:
\begin{eqnarray}
\lefteqn{ A_{ij} \; = \; 
\mathcal{A}_{ij} 
- \tfrac{\lambda}{2} \, I[\mathcal{A}_{kk}] \delta_{ij} 
+ \Bigl\{ \tfrac34 \lambda^2 I^2[\mathcal{A}_{kk}] 
  + \tfrac{\lambda}{2} \, I[\mathcal{A}_{k\ell} \mathcal{A}_{\ell k}] 
  \Bigr\} \delta_{ij} 
  - \tfrac{\lambda}{2} \, I[\mathcal{A}_{kk} \mathcal{A}_{ij} ] } 
\nonumber \\
& & \hspace{-0.6cm} 
- \Bigl\{ \tfrac{9}{8} \lambda^3 I^3[\mathcal{A}_{kk}] 
\!+\! \tfrac{\lambda}{4} \, 
  I^2[3 \mathcal{A}_{k\ell} \mathcal{A}_{\ell k} 
  \!-\! \mathcal{A}_{kk} \mathcal{A}_{\ell\ell} ] 
  \!+\! \tfrac{\lambda^2}{8} ( I[\mathcal{A}_{kk}] )^2 
  \!+\! \tfrac{\lambda}{2} \, I[\mathcal{A}_{k\ell} 
  \mathcal{A}_{\ell m} \mathcal{A}_{mk}] \Bigr\} \delta_{ij} 
\nonumber \\
& & \hspace{-0.6cm} 
+ \tfrac34 \lambda^2 I\Bigl[ I[\mathcal{A}_{kk}] \mathcal{A}_{ij} \Bigr] 
  + \tfrac{\lambda}{2} \, I[\mathcal{A}_{k\ell} 
  \mathcal{A}_{\ell k} \mathcal{A}_{ij}] + \mathcal{O}(\lambda^5) 
\; . \label{Aexpansion}
\end{eqnarray} 

\subsection{The graviton mode function and the tensor power spectrum}

The mode functions $u(t, {\bf k})$ of the graviton field 
$h_{ij}(t, {\bf x})$ are defined in its usual expansion: 
\footnote{By (\ref{ABC}), the relation of $h_{ij}$ to $A_{ij}$ 
is: $h_{ij} = \kappa A_{ij}$.}
\begin{equation}
h_{ij}(t, {\bf x}) =\!
\int_{k=H} \frac{d^3k}{(2\pi)^3}
\Big[ e^{i {\bf k} \cdot {\bf x}} \, u(t, {\bf k}) \, \alpha_{ij}({\bf k})
+ e^{-i {\bf k} \cdot {\bf x}} \, u^*(t, {\bf k}) \, 
\alpha^{\dagger}_{ij}({\bf k}) \Big]
\; . \label{Aij}
\end{equation}
On the other hand, the background free field $\mathcal{A}_{ij}$ 
is a stochastic field which commutes with itself:
\begin{equation}
\mathcal{A}_{ij}(t, {\bf x}) =\!
\int_{k=H}^{H a(t)} \!\! \frac{d^3k}{(2\pi)^3} \;
u_0(t, {\bf k})
\Big[ e^{i {\bf k} \cdot {\bf x}} \, \alpha_{ij}({\bf k})
+ e^{-i {\bf k} \cdot {\bf x}} \, \alpha^{\dagger}_{ij}({\bf k}) \Big]
\; , \label{calAij}
\end{equation}
and its mode functions $u_0(t, {\bf k})$ assume their late time 
limit in the underlying spacetime; for de Sitter that is:
\begin{equation}
u_0(t, {\bf k}) = \frac{H}{\sqrt{2 k^3}}
\; . \label{u0}
\end{equation}
Furthermore, the definition of the tensor power spectrum is:
\begin{equation}
\Delta_h^2(t, {\bf k}) \equiv 
32 \pi G \, \frac{k^3}{2 \pi^2} \, 2 \vert u(t, {\bf k}) \vert^2 
\quad , \quad t \gg t_k
\; , \label{tensor}
\end{equation}
where $t_k$ is the first horizon crossing time.

Given the above, we look for operators proportional to 
$\mathcal{A}_{ij}$; the first order contributions come 
from only a few terms of (\ref{Aexpansion}) so that:
\begin{equation}
\mathcal{A}_{ij} \; \longrightarrow \; 
\mathcal{A}_{ij} 
- \tfrac{\lambda}{2} \, I[\mathcal{A}_{kk} \mathcal{A}_{ij}] 
+ \tfrac34 \lambda^2 I\Bigl[ I[\mathcal{A}_{kk}] \mathcal{A}_{ij} \Bigr] 
+ \tfrac{\lambda}{2} \, I[\mathcal{A}_{k\ell} \mathcal{A}_{\ell k} 
  \mathcal{A}_{ij}] 
+ \dots
\label{firstorder} 
\end{equation}
Hence, the lowest order secular correction comes 
from the vacuum expectation value of the last term 
in (\ref{firstorder}):
\begin{equation}
\mathcal{A}_{ij} \; \longrightarrow \; 
\mathcal{A}_{ij} \, \Bigl\{
1 + \tfrac{\lambda}{2} \, I[ \langle \mathcal{A}_{k\ell} 
\mathcal{A}_{\ell k} \rangle ] + \dots \Bigr\} 
\; . \label{firstcorrection}
\end{equation} 
Because of (\ref{Avev}) we conclude that:
\begin{equation}
\mathcal{A}_{ij} \; \longrightarrow \; 
\mathcal{A}_{ij} \, \Bigl\{
1 + \tfrac{9}{8} \lambda^2 T^2 + \dots \Bigr\}
\; . \label{calAijresult}
\end{equation}
This is a 2-loop effect for the graviton mode function:
\begin{equation}
u(t, {\bf k}) = u_0(t, {\bf k}) 
\Big\{ 1 + \tfrac{9}{8} \lambda^2 T^2 + \dots \Bigr\}
\label{uresult}
\end{equation}
and through (\ref{tensor}) for the tensor power spectrum.

\subsection{The gravitational force}

The gravitational force $F$ is defined in terms of the 
constrained gravitational variable $C$ (\ref{ABC}) in 
the usual way:
\begin{equation}
F(t, r) = - \nabla C(t, r) 
\; . \label{force}
\end{equation}
The relevant gravitational equation appropriate to give 
the response to the presence of a point mass $M$ comes 
from the constraint field equation involving $C$:
\begin{equation}
\widetilde{D}_C \, C 
\, = \, 
\tfrac{\kappa^2 \widetilde{H}^4}{8 \pi^2} a^4 
\sqrt{-\widetilde{g}} \,
\Bigl\{ 2 - 4 \gamma^{ij} A_{ij} \Bigr\} - 4 \pi G M
a \, \delta^3({\bf x}) 
\; , \label{Ceqn}
\end{equation}
where the differential operator $\widetilde{D}_C$ is given 
by \cite{Miao:2025gzm}:
\footnote{The form of the three graviton quadratic operators
$\widetilde{D}_I, \; I={A,B,C}$ can be found in the Appendix.}
\begin{eqnarray}
\lefteqn{\widetilde{D}_C 
\, \equiv \,
\partial_{\mu} \Bigl[ a^2 \sqrt{-\widetilde{g}} 
   \, \widetilde{g}^{\mu\nu} \partial_{\nu} \Bigr] 
- 2 \widetilde{H}^2 a^4 \sqrt{-\widetilde{g}} \; , } 
\label{DC} \\
& & \hspace{-0.7cm} 
= \partial_0 \Bigl[ a^2 N \sqrt{\gamma} \, 
   (\widetilde{g}^{00} \partial_0 \!+\! \widetilde{g}^{0j} \partial_j) \Bigr] 
\!+\! \partial_i \Bigl[ a^2 N \sqrt{\gamma} \, 
   (\widetilde{g}^{i0} \partial_0 \!+\! \widetilde{g}^{ij} \partial_j) \Bigr] 
\!-\! 2 \widetilde{H}^2 a^4 N \sqrt{\gamma} 
\; , \qquad \label{DC2} \\
& & \hspace{-0.7cm} 
= a^2 \partial_i \Bigl[ N \sqrt{\gamma} \, \gamma^{ij} \partial_j \Bigr] 
\!-\! ( \partial_0 \!+\! \partial_i N^i ) \Bigl[
   \tfrac{a^2 \sqrt{\gamma}}{N} ( \partial_0 \!+\! N^j \partial_j ) \Bigr] 
\!-\! 2 \widetilde{H}^2 a^4 N \sqrt{\gamma} 
\; . \label{DC3}
\end{eqnarray}
In the perturbative regime, equation (\ref{DC3}) can be simplified 
by setting the lapse $N = 1$ and the shift $N^i = 0$:
\begin{equation}
\widetilde{D}_C \;\longrightarrow \;
a^2 \partial_i \Bigl[ \sqrt{\gamma} \, \gamma^{ij} \partial_j \Bigr] 
- \partial_0 \Bigl[ a^2 \sqrt{\gamma} \partial_0\Bigr] 
- 2 H^2 a^4 \sqrt{\gamma} 
\; . \label{DC4}
\end{equation}

The induced stress tensor, i.e. the first part of the 
right hand side of (\ref{Ceqn}), can change the homogeneous 
cosmological background but does not affect the response to 
a point source $M$. Moreover, because equation (\ref{Ceqn})
is linear in $C$, we can focus on the solution emanating
solely from the point source term. Therefore the equation 
we must solve is:
\begin{eqnarray}
\widetilde{D}_C \, C
&\!\!\! \equiv \!\!\!&
a^2 \partial_i \Bigl[ \sqrt{\gamma} \, \gamma^{ij} \partial_j C \Bigr] 
- \partial_0 \Bigl[ a^2 \sqrt{\gamma} \, \partial_0 C \Bigr] 
- 2 H^2 a^4 \sqrt{\gamma} \, C 
\nonumber \\
&\!\!\! \simeq \!\!\!& 
- 4 \pi G M a \, \delta^3({\bf x}) 
\; , \label{approxCeqn}
\end{eqnarray}
where we have also incorporated (\ref{DC4}).

Taking into account that the spatial metric $\gamma_{ij} \simeq 
\delta_{ij} \!+\! A_{ij}$, we proceed to expand $\sqrt{\gamma}$ 
and $\gamma^{ij}$:
\begin{eqnarray}
\sqrt{\gamma} &\!\!\! = \!\!\!& 
1 + \tfrac12 A_{kk} + \tfrac18 (A_{kk})^2 
- \tfrac14 A_{k\ell} A_{\ell k} + \dots 
\; , \label{sqrt} \\
\gamma^{ij} &\!\!\! = \!\!\!& 
\delta_{ij} - A_{ij} + A_{ik} A_{kj} + \dots 
\; . \label{inverse}
\end{eqnarray}
It is apparent from (\ref{Aexpansion}) that the vacuum 
expectation value of a single $A_{ij}$ is of order $\lambda^2$
and thus is a 2-loop quantity. Hence the 1-loop contributions 
to $\sqrt{\gamma} \gamma^{ij}$ are:
\begin{eqnarray}
\Bigl\langle \tfrac18 (\widetilde{A}_{kk})^2 
- \tfrac14 \widetilde{A}_{k\ell} \widetilde{A}_{\ell k} 
\Bigr\rangle \, \delta_{ij} 
&\! \longrightarrow \!& 
-\tfrac94 \lambda T \delta_{ij} 
\; , \label{vev1} \\
\Bigl\langle -\tfrac12 \widetilde{A}_{kk} \widetilde{A}_{ij} \Bigr\rangle 
&\! \longrightarrow \!& 
\tfrac32 \lambda T \delta_{ij} 
\; , \label{vev2} \\
\Bigl\langle \widetilde{A}_{ik} \widetilde{A}_{kj} \Bigr\rangle 
&\! \longrightarrow \!& 
\tfrac32 \lambda T \delta_{ij} 
\; , \label{vev3}
\end{eqnarray}
so that the vacuum expectation value of 
$\widetilde{\mathcal{D}}_{C} \, C$ equals:
\begin{eqnarray}
\Bigl\langle \widetilde{D}_C \Bigr\rangle \, C 
&\!\!\! = \!\!\!& 
a^2 \partial_i \Bigl[ \Bigl( 1 + \tfrac34 \lambda + \dots \Bigr) 
   \partial_j C\Bigr]
- \partial_0 \Bigl[ a^2 \Bigl( 1 \!-\! \tfrac94 \lambda T 
   + \dots \Bigr) \partial_0 C \Bigr]
\qquad \nonumber \\
& \mbox{} &  
- 2 H^2 a^4 \Bigl( 1 \!-\! \tfrac94 \lambda T + \dots \Bigr) \, C 
\; . \label{DCvev}
\end{eqnarray}
This allows us to perturbatively solve (\ref{approxCeqn}) 
for the gravitational potential $C(t, r)$ and determine 
the first deviation from its classical value:
\begin{eqnarray}
\Bigl\langle \widetilde{D}_C \Bigr\rangle 
&\!\!\! = \!\!\!&  
\Bigl\langle \widetilde{D}_C \Bigr\rangle \Big\vert_{\lambda^0}
+ \Bigl\langle \widetilde{D}_C \Bigr\rangle \Big\vert_{\lambda^1}
+ \dots
\; , \nonumber \\
C(t, r)
&\!\!\! = \!\!\!&
C(t, r) \Big\vert_{\lambda^0} 
+ C(t, r) \Big\vert_{\lambda^1}
+ \dots
\label{pertsolve}
\end{eqnarray}
By inserting (\ref{pertsolve}) in equation (\ref{approxCeqn})
and solving the equation order by order we obtain the desired
solution for the Newtonian potential displaying an order $G H^2$ 
secular correction:
\begin{equation}
C(t,r) = 
C(t, r) \Big\vert_{\lambda^0} + C(t, r) \Big\vert_{\lambda^1}
=
\tfrac{G M}{a r} \Bigl\{
1 - \tfrac{\kappa^2 H^2}{4 \pi^2} \, \ln[a H r] 
+ \dots \Bigr\} 
\; , \label{Newton}
\end{equation}
where we have expressed the final answer in terms of the
original physical parameters using (\ref{defs}).

\section{Epilogue}

This work was heavily motivated by the need is to demonstrate 
the consistency of the predictions of the proposed LLOG equations 
(\ref{Afinal3}-\ref{Cfinal3}) in their perturbative correspondence 
limit with those of independent explicit perturbative loop 
calculations. Obviously the latter should be carried out with 
the {\it same} gauge condition (\ref{F}) used to derive the LLOG 
equations. In terms of physical quantities, we derived herein 
the perturbative predictions of (\ref{Afinal3}-\ref{Cfinal3}) 
for the graviton mode functions from which the primordial tensor 
power spectrum follows, and for the gravitational force due 
to a point source.

Besides the highly non-trivial check on the LLOG equations 
coming from fully regularized and renormalized perturbative 
loop computations, their ability to penetrate into the
non-perturbative regime should be emphasized. It therefore 
appears that there is a seemingly tractable path to determine 
what happens after perturbation theory breaks down and which 
late-time gravitational cosmological model emerges.

\vspace{0.5cm}

\centerline{\bf Acknowledgements}

This work was partially supported by Taiwan NSTC grants
113-2112-M-006-013 and 114-2112-M-006-020, by NSF grant
PHY-2207514 and by the Institute for Fundamental Theory
at the University of Florida.



\section{Appendix}

\subsection{The ADM decomposition}

The spatial-temporal decomposition of the conformally 
re-scaled metric is \cite{Arnowitt:1962hi}:
\begin{eqnarray}
{\widetilde g}_{\mu\nu} 
&\!\!\!\! = \!\!\!\!&
\begin{pmatrix}
-N^2 \!+\! \gamma_{kl} N^k N^l \;&\; -\gamma_{jl} N^l \\
\\
-\gamma_{ik} N^k & \gamma_{ij} \\
\end{pmatrix}
\label{3+1lower} \\
&\!\!\!\! = \!\!\!\!&
\begin{pmatrix}
\gamma_{kl} N^k N^l \;&\; -\gamma_{jl} N^l \!&\! \\
\\
-\gamma_{ik} N^k \;&\; \gamma_{ij} \\
\end{pmatrix}
-
\begin{pmatrix}
-N \\
\\
0 \\
\end{pmatrix}
_{\!\!\!\mu}
\begin{pmatrix}
-N \\
\\
0 \\
\end{pmatrix}
_{\!\!\!\nu} 
\equiv
{\overline \gamma}_{\mu\nu} \!\!- u_{\mu} u_{\nu}
\; , \qquad \label{3+1lowerB}
\end{eqnarray}
which implies the following form for its inverse: 
\begin{eqnarray}
{\widetilde g}^{\mu\nu}
&\!\!\!\! = \!\!\!\!&
\begin{pmatrix}
-\frac{1}{N^2} & -\frac{N^j}{N^2} \\
\\
-\frac{N^i}{N^2} \;&\; \gamma^{ij} \!-\! \frac{N^i N^j}{N^2} \\
\end{pmatrix}
\label{3+1upper} \\
&\!\!\!\! = \!\!\!\!&
\begin{pmatrix}
0 & 0 \\
\\
0 \;&\; \gamma^{ij} \\
\end{pmatrix}
-
\begin{pmatrix}
\frac{1}{N} \\
\\
\frac{N^i}{N} \\
\end{pmatrix}
^{\!\!\!\mu}
\begin{pmatrix}
\frac{1}{N} \\
\\
\frac{N^j}{N} \\
\end{pmatrix}
^{\!\!\!\nu} 
\equiv
{\overline \gamma}^{\mu\nu} \!\!- u^{\mu} u^{\nu}
\; . \qquad \label{3+1upperB}
\end{eqnarray}
In (\ref{3+1lower}-\ref{3+1upperB}) $N$ is the lapse 
function, $N^i$ is the shift function, and $\gamma_{ij}$ 
is the spatial metric. The ``spatial part'' is
${\overline \gamma}^{\mu\nu}$ and the ``temporal part''
is  $u_{\mu}$.

\subsection{The graviton propagator in $D=4$ dimensions}

\noindent
$\bullet \,$
The graviton propagator is:
\begin{equation}
i [\mbox{}_{\mu\nu} \widetilde{\Delta}_{\rho\sigma}](x;x') 
= 
\sum_{I=A,B,C} [\mbox{}_{\mu\nu} \widetilde{T}^I_{~\rho\sigma}] 
\!\times
i \widetilde{\Delta}_{I}(x;x') 
\; , \label{gravprop} 
\end{equation}
where $\, i \widetilde{\Delta}_{A}(x;x')$ is the massless 
minimally coupled scalar propagator in de Sitter with Hubble 
parameter $\widetilde{H}$ and, where 
$\, i \widetilde{\Delta}_{B}(x;x') \; \& \; 
i \widetilde{\Delta}_{C}(x;x')$ are massive scalar propagators 
in de Sitter which agree only in $D=4$; in that case their mass
is $m^2 = 2\widetilde{H}^2$. The associated tensor factors have 
the remarkable property of being spacetime {\it constants} 
and hence the resulting graviton propagator has excellent 
perturbative calculability:
\begin{eqnarray}
[\mbox{}_{\mu\nu} \widetilde{T}^{A}_{~\rho\sigma}] 
&\!\!\! = \!\!\!& 
2 \,\overline{\gamma}_{\mu (\rho} \overline{\gamma}_{\sigma) \nu} 
- 2 \, \overline{\gamma}_{\mu\nu} \overline{\gamma}_{\rho\sigma} 
\; , \label{T_A} \\
{[} \mbox{}_{\mu\nu} \widetilde{T}^{B}_{~\rho\sigma}] 
&\!\!\! = \!\!\!& 
- 4 \, u_{(\mu} \overline{\gamma}_{\nu) (\rho} u_{\sigma)} 
\; , \label{T_B} \\
{[} \mbox{}_{\mu\nu} \widetilde{T}^{C}_{~\rho\sigma}] 
&\!\!\! = \!\!\!&
\big[ u_{\mu} u_{\nu} + \overline{\gamma}_{\mu\nu} \big] 
\big[ u_{\rho} u_{\sigma} + \overline{\gamma}_{\rho\sigma} \big] 
\; . \label{T_C}
\end{eqnarray}

\noindent
$\bullet \,$
The coincidence limits of the above three scalar 
propagators are:
\begin{eqnarray}
i\widetilde{\Delta}_{A}(x;x')\vert_{x'=x}
&\!\!\! = \!\!\!&
\tfrac{{\widetilde H}^2}{4\pi^2} \ln a + ``\infty"
\; , \label{Axx} \\
i\widetilde{\Delta}_{B}(x;x')\vert_{x'=x}
&\!\!\! = \!\!\!&
-\tfrac{{\widetilde H}^2}{16\pi^2}
\; , \label{Bxx} \\
i\widetilde{\Delta}_{C}(x;x')\vert_{x'=x}
&\!\!\! = \!\!\!&
+\tfrac{{\widetilde H}^2}{16\pi^2}
\; , \label{Cxx}
\end{eqnarray}

\noindent
$\bullet \,$
The ${\widetilde D}_A$ quadratic operator equals:
\begin{equation}
{\widetilde D}_A 
\equiv
\partial_{\alpha} \big[ a^2 {\sqrt {-\widetilde g}} \,
{\widetilde g}^{\alpha\beta} \partial_{\beta} \big]
\; , \label{D_A}
\end{equation}
while the other two quadratic operators are expressible 
in terms of ${\widetilde D}_A$:
\begin{equation}
{\widetilde D}_B 
=
{\widetilde D}_C 
= 
\widetilde{D}_A - 2 \widetilde{H}^2 a^D \sqrt{-\widetilde{g}}
\; . \label{D_B=D_C}
\end{equation}




\begin{thebibliography}{99}

\bibitem{Geshnizjani:2011dk}
G.~Geshnizjani, W.~H.~Kinney and A.~Moradinezhad Dizgah,
JCAP \textbf{11}, 049 (2011)
doi:10.1088/1475-7516/2011/11/049
[arXiv:1107.1241 [astro-ph.CO]].

\bibitem{Ijjas:2013vea}
A.~Ijjas, P.~J.~Steinhardt and A.~Loeb,
Phys. Lett. B \textbf{723}, 261-266 (2013)
doi:10.1016/j.physletb.2013.05.023
[arXiv:1304.2785 [astro-ph.CO]].

\bibitem{Guth:2013sya}
A.~H.~Guth, D.~I.~Kaiser and Y.~Nomura,
Phys. Lett. B \textbf{733}, 112-119 (2014)
doi:10.1016/j.physletb.2014.03.020
[arXiv:1312.7619 [astro-ph.CO]].

\bibitem{Linde:2014nna}
A.~Linde,
doi:10.1093/acprof:oso/9780198728856.003.0006 \hfill \break 
[arXiv:1402.0526 [hep-th]].

\bibitem{Tsamis:1996qm}
N.~C.~Tsamis and R.~P.~Woodard,
Annals Phys. \textbf{253}, 1-54 (1997)
doi:10.1006/aphy.1997.5613
[arXiv:hep-ph/9602316 [hep-ph]].

\bibitem{Tsamis:2011ep}
N.~C.~Tsamis and R.~P.~Woodard,
Int. J. Mod. Phys. D \textbf{20}, 2847-2851 (2011)
doi:10.1142/S0218271811020652
[arXiv:1103.5134 [gr-qc]].


\bibitem{Grishchuk:1977zz}
L.~P.~Grishchuk,
Annals N. Y. Acad. Sci. \textbf{302}, 439 (1977)
doi:10.1111/j.1749-6632.1977.tb37064.x

\bibitem{Myhrvold:1983hx}
N.~P.~Myhrvold,
Phys. Rev. D \textbf{28}, 2439 (1983)
doi:10.1103/PhysRevD.28.2439

\bibitem{Ford:1984hs}
L.~H.~Ford,
Phys. Rev. D \textbf{31}, 710 (1985)
doi:10.1103/PhysRevD.31.710

\bibitem{Allen:1985wd}
B.~Allen and T.~Jacobson,
Commun. Math. Phys. \textbf{103}, 669 (1986)
doi:10.1007/BF01211169

\bibitem{Antoniadis:1986sb}
I.~Antoniadis and E.~Mottola,
J. Math. Phys. \textbf{32}, 1037-1044 (1991)
doi:10.1063/1.529381

\bibitem{Allen:1986tt}
B.~Allen and M.~Turyn,
Nucl. Phys. B \textbf{292}, 813 (1987)
doi:10.1016/0550-3213(87)90672-9

\bibitem{Floratos:1987ek}
E.~G.~Floratos, J.~Iliopoulos and T.~N.~Tomaras,
Phys. Lett. B \textbf{197}, 373-378 (1987)
doi:10.1016/0370-2693(87)90403-5

\bibitem{Higuchi:1991tm}
A.~Higuchi,
Class. Quant. Grav. \textbf{8}, 1983-2004 (1991)
doi:10.1088/0264-9381/8/11/010

\bibitem{Tsamis:1992xa}
N.~C.~Tsamis and R.~P.~Woodard,
Commun. Math. Phys. \textbf{162}, 217-248 (1994)
doi:10.1007/BF02102015

\bibitem{Dolgov:1994cq}
A.~D.~Dolgov, M.~B.~Einhorn and V.~I.~Zakharov,
Phys. Rev. D \textbf{52}, 717-722 (1995)
doi:10.1103/PhysRevD.52.717
[arXiv:gr-qc/9403056 [gr-qc]].

\bibitem{Tsamis:2005hd}
N.~C.~Tsamis and R.~P.~Woodard,
Nucl. Phys. B \textbf{724}, 295-328 (2005)
doi:10.1016/j.nuclphysb.2005.06.031
[arXiv:gr-qc/0505115 [gr-qc]].

\bibitem{Polyakov:2007mm}
A.~M.~Polyakov,
Nucl. Phys. B \textbf{797}, 199-217 (2008)
doi:10.1016/j.nuclphysb.2008.01.002
[arXiv:0709.2899 [hep-th]].

\bibitem{Giddings:2007nu}
S.~B.~Giddings and D.~Marolf,
Phys. Rev. D \textbf{76}, 064023 (2007)
doi:10.1103/PhysRevD.76.064023
[arXiv:0705.1178 [hep-th]].

\bibitem{Perez-Nadal:2007yxe}
G.~Perez-Nadal, A.~Roura and E.~Verdaguer,
Phys. Rev. D \textbf{77}, 124033 (2008)
doi:10.1103/PhysRevD.77.124033
[arXiv:0712.2282 [gr-qc]].

\bibitem{Burgess:2010dd}
C.~P.~Burgess, R.~Holman, L.~Leblond and S.~Shandera,
JCAP \textbf{10} (2010), 017
doi:10.1088/1475-7516/2010/10/017
[arXiv:1005.3551 [hep-th]].

\bibitem{Polyakov:2012uc}
A.~M.~Polyakov,
[arXiv:1209.4135 [hep-th]].

\bibitem{Marolf:2012kh}
D.~Marolf, I.~A.~Morrison and M.~Srednicki,
Class. Quant. Grav. \textbf{30}, 155023 (2013)
doi:10.1088/0264-9381/30/15/155023
[arXiv:1209.6039 [hep-th]].

\bibitem{Anderson:2013ila}
P.~R.~Anderson and E.~Mottola,
Phys. Rev. D \textbf{89}, 104038 (2014)
doi:10.1103/PhysRevD.89.104038
[arXiv:1310.0030 [gr-qc]].

\bibitem{Frob:2013ht}
M.~B.~Fr{\"o}b, D.~B.~Papadopoulos, A.~Roura and E.~Verdaguer,
Phys. Rev. D \textbf{87}, no.6, 064019 (2013)
doi:10.1103/PhysRevD.87.064019
[arXiv:1301.5261 [gr-qc]].

\bibitem{Anninos:2014lwa}
D.~Anninos, T.~Anous, D.~Z.~Freedman and G.~Konstantinidis,
JCAP \textbf{11}, 048 (2015)
doi:10.1088/1475-7516/2015/11/048
[arXiv:1406.5490 [hep-th]].

\bibitem{Frob:2014zka}
M.~B.~Fr{\"o}b, J.~Garriga, S.~Kanno, M.~Sasaki, J.~Soda, 
T.~Tanaka and A.~Vilenkin,
JCAP \textbf{04}, 009 (2014)
doi:10.1088/1475-7516/2014/04/009
[arXiv:1401.4137 [hep-th]].

\bibitem{Dvali:2014gua}
G.~Dvali and C.~Gomez,
Annalen Phys. \textbf{528}, 68-73 (2016)
doi:10.1002/andp.201500216
[arXiv:1412.8077 [hep-th]].

\bibitem{Burgess:2015ajz}
C.~P.~Burgess, R.~Holman and G.~Tasinato,
JHEP \textbf{01}, 153 (2016)
doi:10.1007/JHEP01(2016)153
[arXiv:1512.00169 [gr-qc]].

\bibitem{Brandenberger:2018fdd}
R.~Brandenberger, L.~L.~Graef, G.~Marozzi and G.~P.~Vacca,
Phys. Rev. D \textbf{98} (2018) no.10, 103523
doi:10.1103/PhysRevD.98.103523
[arXiv:1807.07494 [hep-th]].

\bibitem{Baumgart:2019clc}
M.~Baumgart and R.~Sundrum,
JHEP \textbf{07}, 119 (2020)
doi:10.1007/JHEP07(2020)119
[arXiv:1912.09502 [hep-th]].

\bibitem{Brahma:2021mng}
S.~Brahma, A.~Berera and J.~Calder{\'o}n-Figueroa,
Class. Quant. Grav. \textbf{39}, no.24, 245002 (2022)
doi:10.1088/1361-6382/aca066
[arXiv:2107.06910 [hep-th]].

\bibitem{Colas:2022hlq}
T.~Colas, J.~Grain and V.~Vennin,
Eur. Phys. J. C \textbf{82}, no.12, 1085 (2022)
doi:10.1140/epjc/s10052-022-11047-9
[arXiv:2209.01929 [hep-th]].
E as of 02 Nov 2025

\bibitem{Cable:2023gdz}
A.~Cable and A.~Rajantie,
Phys. Rev. D \textbf{109} (2024) no.4, 045017
doi:10.1103/PhysRevD.109.045017
[arXiv:2310.07356 [gr-qc]].

\bibitem{Burgess:2024eng}
C.~P.~Burgess, T.~Colas, R.~Holman, G.~Kaplanek and V.~Vennin,
JCAP \textbf{08}, 042 (2024)
doi:10.1088/1475-7516/2024/08/042
[arXiv:2403.12240 [gr-qc]].

\bibitem{Anninos:2024fty}
D.~Anninos, T.~Anous and A.~Rios Fukelman,
JHEP \textbf{08} (2024), 155
doi:10.1007/JHEP08(2024)155
[arXiv:2403.16166 [hep-th]].

\bibitem{Brahma:2024yor}
S.~Brahma, J.~Calder{\'o}n-Figueroa and X.~Luo,
JCAP \textbf{08}, 019 (2025)
doi:10.1088/1475-7516/2025/08/019
[arXiv:2407.12091 [hep-th]].

\bibitem{Sloth:2025nan}
M.~S.~Sloth,
Phys. Rev. D \textbf{112}, no.2, 2 (2025)
doi:10.1103/bsjy-4tj8
[arXiv:2502.03520 [hep-th]].

\bibitem{Ansari:2025nng}
E.~Ansari, S.~Bhowmick and D.~Ghosh,
[arXiv:2512.11040 [hep-th]].

\bibitem{Ahmadi:2025oon}
Z.~Ahmadi and M.~Noorbala,
[arXiv:2512.17070 [gr-qc]].

\bibitem{Kaplanek:2025moq}
G.~Kaplanek, M.~Mylova and A.~J.~Tolley,
[arXiv:2512.17089 [hep-th]].

\bibitem{Prokopec:2025jrd}
T.~Prokopec,
[arXiv:2512.22958 [gr-qc]].


\bibitem{DHoker:1994rdl}
E.~D'Hoker and S.~Weinberg,
Phys. Rev. D \textbf{50}, R6050-R6053 (1994)
doi:10.1103/PhysRevD.50.R6050
[arXiv:hep-ph/9409402 [hep-ph]].

\bibitem{Donoghue:1993eb}
J.~F.~Donoghue,
Phys. Rev. Lett. \textbf{72}, 2996-2999 (1994)
doi:10.1103/PhysRevLett.72.2996
[arXiv:gr-qc/9310024 [gr-qc]].

\bibitem{Donoghue:1994dn}
J.~F.~Donoghue,
Phys. Rev. D \textbf{50}, 3874-3888 (1994)
doi:10.1103/PhysRevD.50.3874
[arXiv:gr-qc/9405057 [gr-qc]].

\bibitem{Donoghue:2017ovt}
J.~Donoghue,
Scholarpedia \textbf{12}, no.4, 32997 (2017)
doi:10.4249/scholarpedia.32997

\bibitem{Starobinsky:1979ty}
A.~A.~Starobinsky,
JETP Lett. \textbf{30}, 682-685 (1979)

\bibitem{Mukhanov:1981xt}
V.~F.~Mukhanov and G.~V.~Chibisov,
JETP Lett. \textbf{33}, 532-535 (1981)

\bibitem{Mukhanov:1990me}
V.~F.~Mukhanov, H.~A.~Feldman and R.~H.~Brandenberger,
Phys. Rept. \textbf{215}, 203-333 (1992)
doi:10.1016/0370-1573(92)90044-Z

\bibitem{Leonard:2013xsa}
K.~E.~Leonard and R.~P.~Woodard,
Class. Quant. Grav. \textbf{31}, 015010 (2014)
doi:10.1088/0264-9381/31/1/015010
[arXiv:1304.7265 [gr-qc]].

\bibitem{Wang:2014tza}
C.~L.~Wang and R.~P.~Woodard,
Phys. Rev. D \textbf{91}, no.12, 124054 (2015)
doi:10.1103/PhysRevD.91.124054
[arXiv:1408.1448 [gr-qc]].

\bibitem{Glavan:2013jca}
D.~Glavan, S.~P.~Miao, T.~Prokopec and R.~P.~Woodard,
Class. Quant. Grav. \textbf{31}, 175002 (2014)
doi:10.1088/0264-9381/31/17/175002
[arXiv:1308.3453 [gr-qc]].

\bibitem{Radkowski:1970ovx}
A.~F.~Radkowski,
Annals Phys. \textbf{56}, no.2, 319-354 (1970)
doi:10.1016/0003-4916(70)90021-7

\bibitem{Miao:2005am}
S.~P.~Miao and R.~P.~Woodard,
Class. Quant. Grav. \textbf{23}, 1721-1762 (2006)
doi:10.1088/0264-9381/23/5/016
[arXiv:gr-qc/0511140 [gr-qc]].

\bibitem{Miao:2006gj}
S.~P.~Miao and R.~P.~Woodard,
Phys. Rev. D \textbf{74}, 024021 (2006)
doi:10.1103/PhysRevD.74.024021
[arXiv:gr-qc/0603135 [gr-qc]].

\bibitem{Tan:2021ibs}
L.~Tan, N.~C.~Tsamis and R.~P.~Woodard,
Class. Quant. Grav. \textbf{38}, no.14, 145024 (2021)
doi:10.1088/1361-6382/ac0233
[arXiv:2103.08547 [gr-qc]].

\bibitem{Tan:2021lza}
L.~Tan, N.~C.~Tsamis and R.~P.~Woodard,
Phil. Trans. Roy. Soc. Lond. A \textbf{380}, 0187 (2021)
doi:10.1098/rsta.2021.0187
[arXiv:2107.13905 [gr-qc]].

\bibitem{Glavan:2021adm}
D.~Glavan, S.~P.~Miao, T.~Prokopec and R.~P.~Woodard,
JHEP \textbf{03}, 088 (2022)
doi:10.1007/JHEP03(2022)088
[arXiv:2112.00959 [gr-qc]].

\bibitem{Tan:2022xpn}
L.~Tan, N.~C.~Tsamis and R.~P.~Woodard,
Universe \textbf{8}, no.7, 376 (2022)
doi:10.3390/universe8070376
[arXiv:2206.11467 [gr-qc]].

\bibitem{Miao:2021gic}
S.~P.~Miao, N.~C.~Tsamis and R.~P.~Woodard,
JHEP \textbf{03}, 069 (2022)
doi:10.1007/JHEP03(2022)069
[arXiv:2110.08715 [gr-qc]].

\bibitem{Miao:2024nsz}
S.~P.~Miao, N.~C.~Tsamis and R.~P.~Woodard,
Class. Quant. Grav. \textbf{41}, no.21, 215007 (2024)
doi:10.1088/1361-6382/ad7dc8
[arXiv:2405.01024 [gr-qc]].

\bibitem{Starobinsky:1986fx}
A.~A.~Starobinsky,
Lect. Notes Phys. \textbf{246}, 107-126 (1986)
doi:10.1007/3-540-16452-9{\_}6

\bibitem{Miao:2024atw}
S.~P.~Miao, N.~C.~Tsamis and R.~P.~Woodard,
JHEP \textbf{07}, 099 (2024)
doi:10.1007/JHEP07(2024)099
[arXiv:2405.00116 [gr-qc]].

\bibitem{Miao:2024shs}
S.~P.~Miao, N.~C.~Tsamis and R.~P.~Woodard,
Universe \textbf{11}, no.7, 223 (2025)
doi:10.3390/universe11070223
[arXiv:2409.12003 [gr-qc]].

\bibitem{Miao:2025gzm}
S.~P.~Miao, N.~C.~Tsamis and R.~P.~Woodard,
Phys. Rev. D \textbf{112}, no.4, 1 (2025)
doi:10.1103/vbzh-7h69
[arXiv:2507.04308 [gr-qc]].

\bibitem{Miao:2025bmd}
S.~P.~Miao, N.~C.~Tsamis and R.~P.~Woodard,
Phys. Rev. D \textbf{112}, no.12, 126022 (2025)
doi:10.1103/3h7x-d1qw
[arXiv:2508.17787 [gr-qc]].

\bibitem{Arnowitt:1962hi}
R.~L.~Arnowitt, S.~Deser and C.~W.~Misner,
Gen. Rel. Grav. \textbf{40}, 1997-2027 (2008)
doi:10.1007/s10714-008-0661-1
[arXiv:gr-qc/0405109 [gr-qc]].

\bibitem{DeWitt:1960fc}
B.~S.~DeWitt and R.~W.~Brehme,
Annals Phys. \textbf{9}, 220-259 (1960)
doi:10.1016/0003-4916(60)90030-0

\bibitem{Starobinsky:1994bd}
A.~A.~Starobinsky and J.~Yokoyama,
Phys. Rev. D \textbf{50}, 6357-6368 (1994)
doi:10.1103/PhysRevD.50.6357
[arXiv:astro-ph/9407016 [astro-ph]].

\bibitem{Miao:2006pn}
S.~P.~Miao and R.~P.~Woodard,
Phys. Rev. D \textbf{74}, 044019 (2006)
doi:10.1103/PhysRevD.74.044019
[arXiv:gr-qc/0602110 [gr-qc]].

\bibitem{Prokopec:2007ak}
T.~Prokopec, N.~C.~Tsamis and R.~P.~Woodard,
Annals Phys. \textbf{323}, 1324-1360 (2008)
doi:10.1016/j.aop.2007.08.008
[arXiv:0707.0847 [gr-qc]].

\bibitem{Prokopec:2008gw}
T.~Prokopec, N.~C.~Tsamis and R.~P.~Woodard,
Phys. Rev. D \textbf{78}, 043523 (2008)
doi:10.1103/PhysRevD.78.043523
[arXiv:0802.3673 [gr-qc]].








\end{thebibliography}
\end{document}